\documentclass[prd,twocolumn,superscriptaddress,nofootinbib,floatfix,amsmath,amssymb]{revtex4-1}
\usepackage{graphicx}
\usepackage{enumitem}
\usepackage{bm}
\usepackage{rotating}
\usepackage{hyperref}
\usepackage{pbox}
\usepackage{xcolor}
\usepackage{array}
\usepackage{mathtools}
\usepackage{leftidx}
\usepackage{braket}
\usepackage{tensor}
\usepackage{mathrsfs}
\usepackage{float}
\usepackage[normalem]{ulem}

\newcommand{\R}{\mathbb{R}}
\renewcommand{\d}{\text{d}}
\newcommand{\rr}[1]{\left(#1\right)}
\renewcommand{\H}{\hat{H}}

\newcommand{\ii}{\text{i}}
\newcommand{\sx}{\mathsf{x}}
\newcommand{\vp}{{\hat{\phi}}}
\newcommand{\ha}{{\hat{a}}}
\newcommand{\hs}{{\hat{\sigma}}}
\newcommand{\bx}{\bm{x}}

\newcommand{\zm}{\textsc{zm}}
\newcommand{\osc}{\textsc{osc}}
\newcommand{\Z}{\mathbb{Z}}
\DeclareMathOperator{\tr}{\text{Tr}}

\usepackage{tikz}
\usetikzlibrary{arrows.meta,decorations.pathmorphing,math}

\begin{document}

\title{Vacuum entanglement harvesting with a zero mode}

\author{Erickson Tjoa}
\email{e2tjoa@uwaterloo.ca}
\affiliation{Department of Physics and Astronomy, University of Waterloo, Waterloo, Ontario, N2L 3G1, Canada}
\affiliation{Institute for Quantum Computing, University of Waterloo, Waterloo, Ontario, N2L 3G1, Canada}

\author{Eduardo Mart\'in-Mart\'inez,}
\email{emartinmartinez@uwaterloo.ca}
\affiliation{Department of Applied Mathematics, University of Waterloo, Waterloo, Ontario, N2L 3G1, Canada}
\affiliation{Institute for Quantum Computing, University of Waterloo, Waterloo, Ontario, N2L 3G1, Canada}
\affiliation{Perimeter Institute for Theoretical Physics, 31 Caroline St N, Waterloo, Ontario, N2L 2Y5, Canada}

\begin{abstract}
We investigate vacuum entanglement harvesting in the presence of a zero mode. We show that, for a variety of detector models and couplings (namely, Unruh-DeWitt qubit and harmonic oscillator detectors, amplitude and derivative coupling), the results are strongly dependent on the state of the zero mode, revealing an  ambiguity in studies of entanglement harvesting with Neumann or periodic boundary conditions, or in general in spacetimes with toroidal topologies.
\end{abstract}

\maketitle

    \section{Introduction}
    
    Many developments in recent years have shown that fruitful progress in our understanding of quantum field theory (QFT) and fundamental physics can be attained by applying insights and tools from quantum information theory. Much attention has been focused on the entanglement structure in quantum field theory, since it was shown that different regions of QFT vacua contain classical correlations and entanglement even if the regions are spacelike separated \cite{summers1985bell,summers1987bell}. Vacuum entanglements plays a crucial role in a plethora of fundamental phenomena such as the black hole information loss problem  \cite{hawking1975particle,almheiri2013black,marolf2017black, haco2018black}. Furthermore, from the perspective of quantum information, entanglement is a resource that can be used to perform information processing tasks \cite{nielsen2000quantum,chitambar2019quantum,Brandao2015reversibleQRT}, thus being able to extract or use entanglement contained in readily available QFT states would be both of fundamental and practical interest. 
    
    The fact that one can indeed extract entanglement from the vacuum of a QFT  using a pair of initially uncorrelated two-level quantum systems was first established in the pioneering work of Valentini \cite{Valentini1991nonlocalcorr} and later by Reznik et al. \cite{reznik2003entanglement,reznik2005violating}. This phenomenon has gained  attention in recent years and is now known as entanglement harvesting. Entanglement harvesting has been shown to be sensitive to accelerations \cite{salton2015acceleration}, time dependence of the interaction and number of spacetime dimensions \cite{pozas2015harvesting}, spacetime curvature \cite{Steeg2009,kukita2017harvesting,henderson2018harvesting,ng2018AdS}, spacetime topology \cite{smith2016topology}, boundary conditions \cite{henderson2019entangling,cong2019entanglement}, field state \cite{simidzija2018harvesting}, and more recently also indefinite causal order \cite{Henderson2020temporal}. In more practical settings, entanglement harvesting has  been investigated in different realistic experimental setups \cite{olson2011entanglement,olson2012extraction,sabin2010dynamics,sabin2012extracting,EMM2013farming,ardenghi2018entanglement,beny2018energy}. To study this phenomenon it is common to consider the Unruh-DeWitt (UDW) particle detector model \cite{Unruh1979evaporation,DeWitt1979}---which consists of the monopole coupling of a detector to a scalar field---is a particularly natural simplified model of light-matter interaction. Studies of entanglement harvesting with the more complex full-fledged electromagnetic field interaction and realistic atomic models have indeed revealed that the fundamental features of this interaction were captured already at the level of the UDW model \cite{pozas2016entanglement}.  
    
    A massless quantum scalar field in flat spacetime subject to periodic or Neumann boundary conditions in all spatial directions is known to exhibit a zero mode \cite{Tjoa2019zeromode,EMM2014zeromode}, which is associated to the spatially constant Fourier component of the field's mode decomposition. It also arises for spacetimes with compact spatial section like de Sitter when the field is minimally coupled to curvature \cite{Yazdi2017, Page2012deSitter}, and also for non-scalar fields such as Dirac fermions with analogous boundary conditions \cite{Louko2016fermionZM,louko2016fermionspin}. Since zero modes do not admit a Fock representation, several studies have opted to remove them by hand, which is akin to assuming that it has a negligible impact on the physics in question (e.g., among others, \cite{Robles2017thermometryQFT,Brown2013Amplification,Lorek2014tripartite,Hummer2016bosonfermionZM,Brenna2016antiUnruh}). Indeed in some cases the effect of the zero mode can be minimized or neglected \cite{EMM2014zeromode}, but in general they have strong phenomenological consequences in the light-matter interaction, since it affects particle detectors' responses, the field stress-energy tensor, and two-detector entanglement dynamics \cite{EMM2014zeromode,Lin2016entangleCylin}. Perhaps more importantly, ignoring the zero mode yields strong causality violations in particle detector models \cite{Tjoa2019zeromode}.
    
    In this paper we study the role of the zero mode on the entanglement dynamics of two Unruh-DeWitt detectors on $(1+1)$-dimensional Einstein cylinder. We consider both  the case of  two-qubits and the case of two harmonic oscillators as detector models, and both in the case of the usual UDW coupling and the less common (but employed in \cite{Lin2016entangleCylin}) derivative coupling \cite{Louko2014firewall,Raval1996stochastic,Davies2002negativeenergy,Aubry2014derivative,Wang2014mirror}. We will see that, in all cases, the zero mode contribution can never be ignored when considering the entanglement dynamics of the field. In particular we show that even when choosing states of minimal uncertainty for the zero mode, the impact cannot be made zero in general. Furthermore there is an important dependence on the state of the zero mode in the entanglement dynamics of the detectors. Because of this, we will finally argue that there is no \textit{a priori} way to choose the state of the zero mode and therefore the study of entanglement harvesting (and even detector dynamics) in the presence of a zero mode when the results depend on its state produces ambiguous results.

    This paper is organized as follows. In Section~\ref{sec: setup} we introduce the Unruh-DeWitt formalism for entanglement harvesting and a brief review of the zero mode. In Section~\ref{sec: results} we show the impact of zero mode on the entanglement negativity of the two detectors' joint density matrix. In Section~\ref{sec: othermodels} we briefly discuss how the results generalize to other UDW-like models such as derivative coupling model, harmonic oscillator detectors, and in higher dimensions. 
    We make our conclusions in Section~\ref{sec: conclusion}. We adopt the units $c=\hbar=1$ and also positive signature for the spacetime metric. We will also sometimes use $\sx\equiv (t,x)$ to denote a spacetime point.
    
    \section{Setup}
    \label{sec: setup}
    
    \subsection{Scalar field in Einstein cylinder}
    
    
    We consider a quantum massless scalar field $\vp(t,x)$ in a (1+1)-dimensional Einstein cylinder, i.e. flat spacetime with topological identification $(t,x)\sim (t,x+L)$ such that the spacetime is topologically $\R\times S^1$. This is equivalent to applying periodic boundary conditions on the scalar field. The quantization inertial frame of coordinates $\sx=(t,x)$ will be called  `laboratory frame' throughout the paper. In the lab frame, the field admits a mode decomposition given by
    \begin{align}
        \vp(t,x) &= \vp_{\zm}(t) + \vp_\osc(t,x)\,,\\
        \vp_\osc (t,x)&=  \sum_{n\neq 0}\frac{1}{\sqrt{4\pi |n|}}\left[\ha_n e^{-\ii |k_n| t+\ii k_n x}+\text{h.c.}\right]\,,
    \end{align}
    where $k_n = 2\pi n/L$ and $n\in \Z\setminus\{0\}$. Notice that for periodic boundary conditions there is a zero mode, that is a spatially constant mode $\vp_\zm(t)$. The modes in $\vp_\osc(t,x)$ will be referred to as the `oscillator modes'. 
    
    In the context of light-matter interaction, particle detectors such as those in the Unruh-DeWitt model necessarily couple to the zero mode. This coupling to the zero mode is essential. Indeed, if one ignores the coupling of the detector to the zero mode one would get physically unacceptable behaviour in the detector-field interaction. Namely, the detector model critically violates causality and can signal faster than light if the zero mode is ignored \cite{Tjoa2019zeromode}. What is more, once we consider that the zero mode needs to be part of the light-matter coupling model, one cannot just assume its impact on the dynamics is negligible: it has been shown that the zero mode may induce non-trivial effects on the physics of particles detectors \cite{EMM2014zeromode,Louko2016fermionZM,louko2016fermionspin}.
    
    For $\hat\phi_{\osc}$ we can define a Fock vacuum $\ket{0}$ as the state satisfying $\hat a_n\ket{0}=0$ for all $n\neq 0$ with the usual canonical commutation relation  $[\ha_i,\ha^\dagger_j] = \delta_{ij}\openone$. However, the zero mode does not admit a Fock vacuum since its dynamical behaviour is the same as that of a quantized non-relativistic free particle \cite{EMM2014zeromode,Tjoa2019zeromode}. Indeed, the classical Lagrangian of the zero mode can be written as
    \begin{align}
        \mathcal{L}_\zm = \frac{L\dot{Q}^2}{2}\,,
    \end{align}
    where $Q$ is the spatially constant Fourier component of classical field $\phi$, i.e. $Q=\phi_{\zm}(t)$. Therefore, the zero mode is dynamically equivalent to a non-relativistic free particle on the real line $\R$ with effective mass $L$, and the field amplitude $\hat\phi_{\zm}$ can be viewed as a generalized position variable in phase space. Defining $P$ to be the conjugate momentum to $Q$, the classical Hamiltonian reads
    \begin{align}
        H_\zm = \frac{P^2}{2L}\,,\hspace{0.5cm} P = \frac{\partial \mathcal{L}_\zm}{\partial \dot{Q}}\,,
    \end{align}
    Quantization of the zero mode then proceeds the same way as the quantization of a non-relativistic free particle, where we promote the classical variables $Q,P$ to operators in Heisenberg picture $\hat Q,\hat P$ respectively. If we now define $\hat Q_S,\hat P_S$ to be the corresponding operators in the Schr\"odinger picture with canonical commutation relation $[\hat Q_S,\hat P_S] = \ii\openone$, it follows that $\hat\phi_{\zm}(t)$ can be expressed as
    \begin{align}
        \hat \phi_{\zm}(t) = \hat Q(t) = \hat Q_S+\frac{\hat P_St}{L}\,.
    \end{align}
    We note that the construction of the zero mode as done above can be generalized to arbitrary $(n+1)$ dimensional spacetime $M$ where the spacelike hypersurface is topologically an $n$-torus ($M\cong \R\times \mathbb{T}^n$), where $\mathbb{T}^n = S^1\times...\times S^1$ \cite{Tjoa2019zeromode}.

    \subsection{Unruh-DeWitt model}

    We consider two spacelike-separated observers, Alice and Bob, each carrying one UDW detector that can interact with the field locally.  In our setup we consider each detector to be a two-level system co-moving relative to the quantization frame whose coordinates are $(t,\bm{x})$, so that the proper time of each detector is $\tau=t$ (see e.g., \cite{Pablo2018rqo}). The monopole moment of each detector in the interaction picture is given by \cite{DeWitt1979}
    \begin{align}
        \hat\mu_\nu(t) = \hs^+_\nu e^{\ii\Omega_\nu t }+\hs^-_\nu e^{-\ii\Omega_\nu t}\,,
    \end{align}
    where $\nu=\{\text{A},\text{B}\}$ denotes Alice and Bob respectively and $\Omega_\nu$ is the energy gap of detector $\nu$. We denote $\ket{g_\nu},\ket{e_\nu}$ to be the ground and excited states of the detectors respectively, the $\mathfrak{su}(2)$ ladder operators are simply $\hs^+_\nu=\ket{e_\nu}\!\bra{g_\nu}$ and $\hs^-_\nu=\ket{g_\nu}\!\bra{e_\nu}$.
    
    The interaction Hamiltonian of the detectors-field system in $(n+1)$-dimensional spacetime is given is given by
    \begin{align}
        \H_I(t) &= \sum_{\nu=\{\textsc{a,b}\}}\lambda_\nu\chi_\nu(t)\hat\mu_\nu(t)\int\d^n\bx\, F_\nu(\bx-\bx_\nu)\vp\left(t,\bx\right)\,,
    \end{align}
    where $F_\nu(\bx)$ is the spatial smearing of detector $\nu$, which is centered at $\bx_\nu$, $\chi_\nu(t)$ is the switching function and $\lambda_\nu$ the coupling strengths. The monopole moment operators in this Hamiltonian are understood as $\hat\mu_\textsc{a}(t) \equiv \hat\mu_\textsc{a}(t)\otimes \openone_\textsc{b}$ and $\hat\mu_\textsc{b}(t) \equiv \openone_\textsc{a}\otimes \hat\mu_\textsc{b}(t)$.

    We assume that the system is initialized in the following uncorrelated state
    \begin{align}
        \hat\rho_0 = \ket{g_\textsc{a}}\!\bra{g_\textsc{a}}\otimes\ket{g_\textsc{b}}\!\bra{g_\textsc{b}}\otimes\hat\rho_{\phi}\,,
    \end{align}
    where $\hat\rho_{\phi}$ is the field state. The field state can be further decomposed into the zero mode and oscillator mode components. For the purposes of this paper, we assume that the field starts in an initially uncorrelated product state of the zero mode and the oscillator mode $\hat\rho_\phi = \hat\rho_\zm\otimes\hat\rho_\osc$, and for simplicity assume that $\hat\rho_\osc$ has vanishing one-point function, i.e.
    \begin{align}
        \tr(\hat\rho_\osc\vp_\osc(\sx))=0\,.    
        \label{eq: vanishing-one-point}
    \end{align}
    This assumption is still very general and includes, e.g., the vacuum state of the field, squeezed vacuum or thermal states of any temperature (any Gaussian state whose Wigner function is centered at zero).  
    
    The time evolution operator is given by the usual time-ordered exponential
    \begin{align}
        \hat U &= \mathcal{T}\exp \left(-\int_{-\infty}^\infty \d t\,\H_I(t)\right)\,.
    \end{align} 
    Working perturbatively up to second order in $\lambda$, we can write
      \begin{equation}
            \hat U = \hat U^{(0)}+\hat U^{(1)} + \hat U^{(2)} + O(\lambda^3)\,,
        \end{equation}
    where $\hat U^{(j)}$ is of order $\lambda^j$. Up to second order, these are given by
    \begin{subequations}
        \begin{align}
            \hat U^{(0)} &= \openone\,,\\
            \hat U^{(1)} &= -\ii\int_{-\infty}^\infty\d t\,\hat H_I(t)\,,\\
            \hat U^{(2)} &= -\int_{-\infty}^\infty\d t\int_{-\infty}^t\d t'\,\hat H_I(t)\hat H_I(t')\,.
        \end{align}
    \end{subequations}
    Therefore, we can write the time-evolved state as
    \begin{align}
        \hat\rho = \hat U\hat\rho_0\hat U^\dagger = \sum_{j=1}^\infty\hat\rho^{(j)}\,,
    \end{align} 
    where $\hat\rho^{(j)}$ is of order $\lambda^j$ in perturbation theory. Again, to second order, the perturbative contributions to time evolution are given by
    \begin{subequations}
        \begin{align}
        \hat\rho^{(0)} &= \hat\rho_0\,,\\
        \hat \rho^{(1)} &= \hat U^{(1)}\hat\rho_0 + \hat\rho_0 \hat U^{(1)\dagger}\,,\\
        \hat \rho^{(2)} &= \hat U^{(2)}\hat\rho_0 + \hat U^{(1)}\hat\rho_0 \hat U^{(1)\dagger} +  \hat\rho_0 \hat U^{(2)\dagger}\,.
        \end{align}
    \end{subequations}
    
    By tracing out the field, the joint reduced density matrix for the two detectors (after interaction) in the basis $\{\ket{g_\textsc{a}g_\textsc{b}},\ket{g_\textsc{a}e_\textsc{b}},\ket{e_\textsc{a}g_\textsc{b}},\ket{e_\textsc{a}e_\textsc{b}}\}$ reads (see e.g, \cite{pozas2015harvesting})
    \begin{align}
        \hat\rho = \begin{pmatrix}
        1-\mathcal{L}_{\textsc{aa}}-\mathcal{L}_{\textsc{bb}} & 0 & 0 & \mathcal{M}^*\\
        0 & \mathcal{L}_{\textsc{bb}} & \mathcal{L}_{\textsc{ba}}   & 0 \\
        0 & \mathcal{L}_{\textsc{ab}} & \mathcal{L}_{\textsc{aa}} & 0\\
        \mathcal{M} & 0   & 0   & 0
        \end{pmatrix}+O(\lambda^3)\,,
    \end{align}
    where 
    \begin{widetext}
    \begin{equation}
        \begin{aligned}
            \mathcal{L}_{ij} &= \lambda_i\lambda_j\int \d t\d t'\,e^{-\ii\Omega_it}e^{\ii\Omega_jt'}\chi_i(t)\chi_j(t')W(\sx_i(t),\sx_j(t'))\,,\hspace{0.5cm} i,j = \textsc{A},\textsc{B}\,,\\
        \mathcal{M}
        &= -\lambda_\textsc{a}\lambda_\textsc{b}\int_{-\infty}^\infty\d t\int^\infty_{-\infty}\d t'\chi_\textsc{a}(t)\chi_\textsc{b}(t')e^{\ii\Omega_\textsc{a} t +\ii\Omega_\textsc{b}t'}\bigr(\Theta(t-t')W(\sx_\textsc{a}(t),\sx_\textsc{b}(t'))+\Theta(t'-t)W(\sx_\textsc{b}(t'),\sx_\textsc{a}(t))\bigr)\,.
        \label{eq: densitymatrixelements}
        \end{aligned}
    \end{equation}
    \end{widetext}
     $W(\sx_\textsc{a}(t),\sx_\textsc{b}(t'))$ is the pullback of the Wightman function along the detectors' trajectory, i.e.
    \begin{align}
        W(\sx_\textsc{a}(t),\sx_{\textsc{b}}(t')) = \tr(\hat \rho_{\phi}\vp(\sx_\textsc{a}(t))\vp(\sx_\textsc{b}(t')))\,.
    \end{align}
    We note that, for our case, the full Wightman function can be conveniently split into two parts, namely the zero mode and the oscillator mode Wightman functions:
    \begin{align}
        W(\sx,\sx') &= W_{\zm}(t,t')+W_{\osc}(\sx,\sx')\,,\\
        W_{\zm}(t,t') &= \tr(\hat\rho_\zm\hat\phi_\zm(t)\hat\phi_\zm(t'))\,,\\
        W_{\osc}(\sx,\sx') &= \tr(\hat\rho_\osc\hat\phi_\osc(\sx)\hat\phi_\osc(\sx'))\,.
    \end{align}
    To see this, observe that
    \begin{align}
        &W(\sx,\sx') \notag\\
        &= \tr_\phi\rr{\hat\rho_0 \vp(\sx)\vp(\sx')}\notag \\
        &= 
        \tr_\phi\rr{\hat\rho_0 \vp_{\zm}(\sx)\vp_{\zm}(\sx')}+
        \tr_\phi\rr{\hat\rho_0\vp_{\zm}(\sx)\vp_{\osc}(\sx')}\notag\\
        &\hspace{0.4cm}+
        \tr_\phi\rr{\hat\rho_0\vp_{\osc}(\sx)\vp_{\zm}(\sx')}+
        \tr_\phi\rr{\hat\rho_0 \vp_{\osc}(\sx)\vp_{\osc}(\sx')}\notag\\
        &= 
        \tr\rr{\hat\rho_\zm\vp_{\zm}(\sx)\vp_{\zm}(\sx')}+
        \tr\rr{\hat\rho_\osc \vp_{\osc}(\sx)\vp_{\osc}(\sx')}
       \notag\\
        &\hspace{0.4cm}+ \tr\rr{\hat\rho_\zm\vp_{\zm}(\sx)}\tr\rr{\hat\rho_\osc\vp_{\osc}(\sx')}\notag\\
        &\hspace{0.4cm}+
        \tr\rr{\hat\rho_\osc\vp_{\osc}(\sx')}\tr\rr{\hat\rho_\zm\vp_{\zm}(\sx)}\notag\\
        &=  W_{\zm}(t,t')+W_{\osc}(\sx,\sx')\,,
    \end{align}
    where in the last equality we have used the vanishing one-point function \eqref{eq: vanishing-one-point}. This decomposition simplifies the analysis as it allows us to cleanly separate the contribution coming from the zero mode and the oscillatory modes.
    For example, in the reduced detector density matrix \eqref{eq: densitymatrixelements} we have that
    $\mathcal{L}_{jj} = \mathcal{L}_{jj,\zm}+\mathcal{L}_{jj,\osc}$ and $\mathcal{M}= \mathcal{M}_{\zm}+\mathcal{M}_{\osc}$ as they are all linear in the full Wightman function $W(\sx,\sx')$. 
    
    Finally, any Wightman function can be written as the sum of the Hadamard part (anti-commutator) $C^+$ and the Pauli-Jordan part (commutator) $C^-$:
    \begin{align}
        W_{\osc/\zm}(\sx,\sx') &= \frac{1}{2}C^+_{\osc/\zm}(\sx,\sx') + \frac{1}{2}C^-_{\osc/\zm}(\sx,\sx')\,,
    \end{align}
    where
    \begin{align}
        &C^+_{\osc/\zm}(\sx,\sx')\notag\\
        &= \tr\rr{\hat\rho_{{\osc/\zm}}\left\{\vp_{\osc/\zm}(\sx),\vp_{\osc/\zm}(\sx')\right\}} \,,\\
        &C^-_{\osc/\zm}(\sx,\sx') \notag\\
        &= \tr\rr{\hat\rho_{\osc/\zm}\left[\vp_{\osc/\zm}(\sx),\vp_{\osc/\zm}(\sx')\right]} \,.
    \end{align}
    The Hadamard functions are state-dependent while the commutator expectations are independent of the field state chosen. We are going to focus on vacuum entanglement harvesting, so, for our purposes we will consider the field to be in a Fock vacuum $\hat\rho_{\osc} = \ket{0}\!\!\bra{0}_{\osc}$, which certainly satisfies Eq.~\eqref{eq: vanishing-one-point}. The Hadamard functions read
    \begin{align}
        &C^+_{\osc}(\sx,\sx')  \notag\\
        &= -\frac{\log \left(1-e^{\frac{2 i \pi  (\Delta u-i\epsilon)}{L}}\right)}{4 \pi } - \frac{\log \left(1-e^{\frac{2 i \pi  (\Delta v-i\epsilon)}{L}}\right)}{4 \pi }\notag\\
        &\hspace{.4cm}-\frac{\log \left(1-e^{-\frac{2 i \pi  (\Delta u - i\epsilon)}{L}}\right)}{4 \pi }-\frac{\log \left(1-e^{-\frac{2 i \pi  (\Delta v-i\epsilon)}{L}}\right)}{4 \pi }\,,\\
        &C^+_{\zm}(t,t') = 2\braket{Q_S^2} +
        2\braket{P_S^2}\frac{tt'}{L^2} +  \braket{\{Q_S,P_S\}}\frac{t+t'}{L} \,,
        \label{eq: hadzm}
    \end{align}
    while the commutators are
    \begin{align}
        &C^-_{\osc}(\sx,\sx') \notag\\
        &= -\frac{\log \left(1-e^{\frac{-2 i \pi  (\Delta u-i\epsilon)}{L}}\right)}{4 \pi }-\frac{\log \left(1-e^{\frac{-2 i \pi  (\Delta v-i\epsilon)}{L}}\right)}{4 \pi }\notag\\
        &\hspace{0.4cm}+\frac{\log \left(1-e^{\frac{2 i \pi  (\Delta u - i\epsilon)}{L}}\right)}{4 \pi }+\frac{\log \left(1-e^{\frac{2 i \pi  (\Delta v-i\epsilon)}{L}}\right)}{4 \pi }\,,\\
        &C^-_{\zm}(t,t') =  -\frac{i\Delta t}{L}\,.
        \label{eq: comzm}
    \end{align}
    Here we use the shorthand $\Delta t = t-t'$, $\Delta u = u-u'$, $\Delta v = v-v'$, where $u = t-x$, $v=t+x$. Since we want to prove that entanglement harvesting is indeed strongly dependent on the choice of zero mode state, let us make a similar choice as in \cite{EMM2014zeromode} and consider the zero mode in what would be the ground state of a quantum harmonic oscillator with the following first and second moments:
    \begin{equation}
    \begin{aligned}
        \braket{Q_S} &=  \braket{P_S} = \braket{\{Q_S,P_S\}} = 0 \,,\\
        \braket{Q_S^2} &= \frac{1}{2\gamma}\,,\hspace{0.5cm}    \braket{P_S^2} = \frac{\gamma}{2}\,.
        \label{eq: moments}
    \end{aligned}
    \end{equation}
    This state has the property that it saturates the Heisenberg uncertainty relation and strongly simplify all our subsequent calculations. The constant $\gamma$ that appears above is related to the mass $m$ and frequency $\omega$ of a standard quantum harmonic oscillator by $\gamma=m\omega/(2\hbar)$. For this choice, the Hadamard piece for the zero mode reads
    \begin{align}
        C^+_{\zm}(t,t') &= \frac{1}{\gamma} + \gamma\frac{ tt'}{L^2}\,.
    \end{align}

    \section{Results and discussion}
    \label{sec: results}
    
    In order to measure the amount of entanglement between the detectors after interaction with the field, we employ the negativity $\mathcal{N}$ \cite{vidal2002computable}, which for two two-dimensional systems (like our two-detector system) is an entanglement measure. Up to second-order in perturbation theory, due to translational invariance this boils down to a very simple calculation involving the matrix elements of \eqref{eq: densitymatrixelements}:
    \begin{align}
        \mathcal{N} = \max\{0,|\mathcal{M}| - \mathcal{L}_{jj}\} + O(\lambda^3)\,,\hspace{0.5cm} j = \textsc{A},\textsc{B}\,.
        \label{eq: negativity}
    \end{align}
    This expression has the advantage that it cleanly separates the local noise terms $\mathcal{L}_{jj}$ coming from each detector's response and the non-local term $\mathcal{M}$, thus entanglement is generated between the two detectors if the non-local term dominates local noise. For two qubits, negativity vanishes if and only if the two-qubit state is separable \cite{peres1996separability,Horodecki996separable,vidal2002computable}. 
    
    We will now explicitly compute the matrix elements $\mathcal{M}$ and $\mathcal{L}_{jj}$. For simplicity we assume that $\Omega_\textsc{a}=\Omega_\textsc{b}=\Omega$, $\lambda_\textsc{a}=\lambda_\textsc{b}=\lambda$, and that the switching functions  are Gaussians \footnote{We use $e^{-t^2/2T^2}$ instead of the more commonly used $e^{-t^2/T^2}$ in UDW literature because the expressions for the zero mode contributions to the detector density matrix is cleaner (without factors of $1/4$ around).}
    \begin{align}
        \chi_\textsc{a}(t)=\chi_\textsc{b}(t) \eqqcolon \chi(t) = e^{-\frac{t^2}{2T^2}}\,.
        \label{eq: gaussian}
    \end{align} 
    We will also focus on the pointlike detector limit where the smearing function $F(x) = \delta(x)$. Note that spatial smearing does not impact the zero mode at all because the Wightman function for the zero mode is independent of spatial coordinates, thus this simplification will not affect the physics of the zero mode. Note that due to Gaussian tails of the switching, the observers will only be approximately space-like but this will not affect the main claims of our results.
    
    For the oscillator part, we evaluate $\mathcal{M}_\osc$ and $P_\osc$ numerically. For the zero mode, the contribution to the joint detector density matrix elements admits closed expressions via Fourier transform and convolution theorem. 
    We first note that in the simplest case where the detectors are at rest relative to the quantization frame, we have $\mathcal{L}_{\textsc{aa},\zm}=\mathcal{L}_{\textsc{bb},\zm} \eqqcolon \mathcal{L}_{\zm}$, since the Wightman function of the zero mode is  invariant under spatial translations. Therefore, we can write
    \begin{align}
        \mathcal{L}_{\zm}   &= \frac{1}{2}\rr{\mathcal{L}_\zm^++\mathcal{L}_\zm^-}\,,
    \end{align}
    where 
    \begin{widetext}
    \begin{align}
        \mathcal{L}_\zm^+ &= \lambda^2\int \d t\,\d t'\chi(t)\chi(t')e^{-\ii\Omega(t-t')}C^+_\zm(t,t') = \lambda^2\frac{2 \pi  T ^2 e^{-T ^2 \Omega ^2} \left(\gamma ^2 T ^4 \Omega ^2+L^2\right)}{\gamma  L^2}\,,\\
        \mathcal{L}_\zm^- &= \lambda^2\int \d t\,\d         t'\chi(t)\chi(t')e^{-\ii\Omega(t-t')}C^-_\zm(t,t') = 
        -\lambda^2\frac{4 \pi  T ^4 \Omega  e^{-T ^2 \Omega ^2}}{L}\,.
    \end{align}
    \end{widetext}
    Therefore, the zero mode contribution reads
    \begin{align}
        \mathcal{L}_\zm = \lambda^2\frac{\pi  e^{-T ^2 \Omega ^2} \left(L T -\gamma  T ^3 \Omega \right)^2}{\gamma  L^2}\,.
        \label{eq: Lzm}
    \end{align}
    Note that this term does not vanish as $L\to \infty$, and instead it tends to a finite positive value
    \begin{align}
        \lim_{L\to\infty} \mathcal{L}_\zm = \lambda^2\frac{\pi  T ^2 e^{-T ^2 \Omega ^2}}{\gamma}\,,
    \end{align}
    thus we do \textit{not} recover the free-space limit where the zero mode does not exist. This is particularly remarkable since, for the response of particle detectors, it implies that the limit of very large cavity does not commute with the quantization of the field. That is, in the presence of detectors, a very large periodic cavity is not the same as Minkowski space. From an operational perspective this means that a detector that is switched on for a finite time can tell that it lives in a periodic cavity even if very far away from the `walls' through the zero mode contributions to its transition probability. As we will see later this is a result of the UDW coupling, and this phenomenon is not there if we consider derivative coupling, highlighting, as already discussed in \cite{EMM2014zeromode} that the derivative coupling suffers less from the zero-mode ambiguities than the usual UDW. We will hold further discussion of this until Section~\ref{sec: results}. We also note that $\mathcal{L}_\zm\to 0$ as $T\to \infty$, i.e. the zero mode contribution to the transition probability vanishes in the long interaction limit.
    
    Note that we can also split the non-local term into the commutator and anti-commutator pieces,
    \begin{align}
        \mathcal{M}_{\zm}   &= \frac{1}{2}\rr{\mathcal{M}_\zm^++\mathcal{M}_\zm^-}\,.
    \end{align}
    Closed form expressions can be found as well from Eq.~\eqref{eq: densitymatrixelements}:
    \begin{align}
        \mathcal{M}^+_\zm &= \lambda^2\frac{4 \ii \sqrt{\pi } T^3 e^{-\Omega^2 T^2}}{L} \,,\\
        \mathcal{M}^-_\zm &= -\lambda^2\frac{2\pi  T ^2 e^{-T ^2 \Omega ^2} \left(L^2-\gamma ^2 T ^4 \Omega ^2\right)}{\gamma  L^2}\,,
    \end{align}
    and together they give
    \begin{align}
        \mathcal{M}_\zm = -\lambda^2\frac{e^{-T^2 \Omega ^2} \left(\pi  L^2 T^2-2 \ii \sqrt{\pi } \gamma  L T^3-\pi  \gamma ^2 T^6 \Omega ^2\right)}{\gamma  L^2}
        \label{eq: Mzm}
    \end{align}
    These also do not vanish in the $L\to\infty$ limit, and curiously this limit is \textit{equal} to the large cavity limit of $\mathcal{L}_\zm$ (up to a sign):
    \begin{align}
        \lim_{L\to\infty} \mathcal{M}_\zm = -\lambda^2\frac{\pi  T^2 e^{-\Omega^2 T^2}}{\gamma} = 
        -\lim_{L\to\infty} \mathcal{L}_{jj,\zm}\,.
        \label{eq: samelimit}
    \end{align}
    Again, we note that as $L\to\infty$, the zero mode contribution does not vanish for any finite interaction time $T$. This  means that the procedure of coupling detector to a field \textit{does not commute} with the large cavity limit. This difference between cavity and continuum UDW model was also observed in the study of resonance in UDW model \cite{Tjoa2020resonance}. Using Eq.~\eqref{eq: negativity}, we get
    \begin{align}
        \mathcal{N} = \max\left\{ 0, (|\mathcal{M}_\osc+\mathcal{M}_\zm|-\mathcal{L}_\osc-\mathcal{L}_\zm) \right\} + O(\lambda^4)\,.
        \label{eq: negativity-large-L-limit}
    \end{align}
    Eqs.~\eqref{eq: Lzm} and \eqref{eq: Mzm} make clear that the zero mode contribution to negativity vanishes at second order in perturbation theory when either $T\to\infty$ (i.e., the long time limit) or $\gamma\to\infty$ (i.e. the Gaussian state is infinitely squeezed into the eigenstate of $Q_S$ which has infinite $\braket{P^2_S}$).

        \begin{figure*}[tp]
            \includegraphics[scale=0.8]{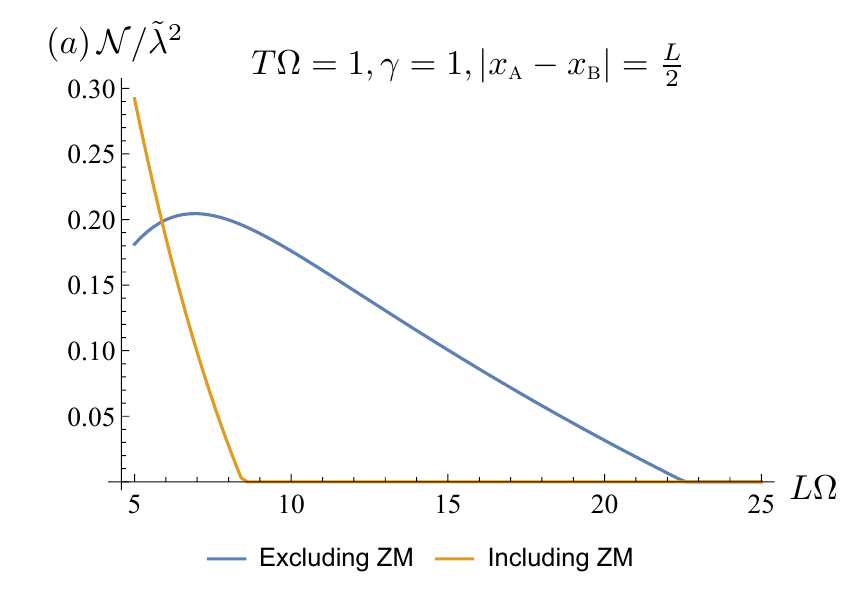}
            \includegraphics[scale=0.8]{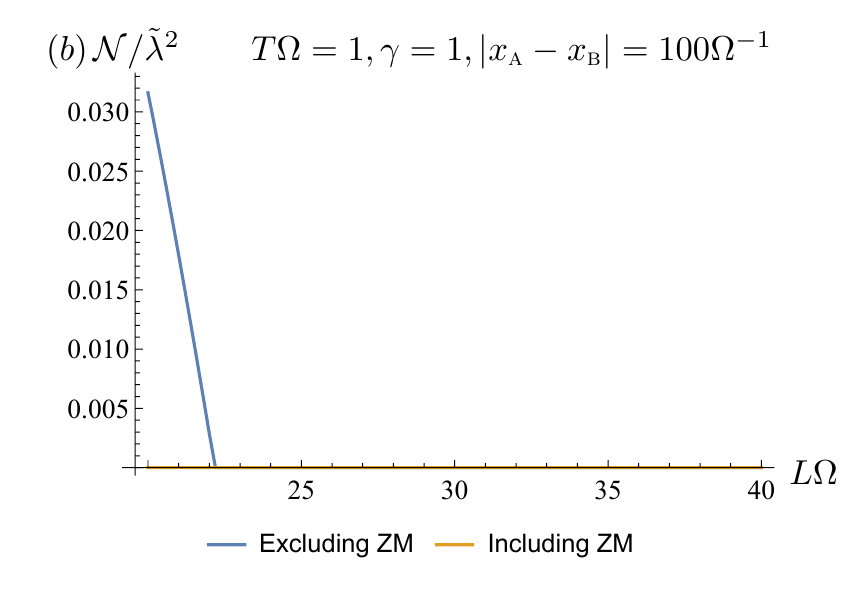}
            \caption{Negativity as a function of the circumference of the cylinder $L$ in units of $\Omega$, for $\gamma=1$ and $T\Omega=1$. Here $\tilde\lambda = \lambda\Omega^{(n-3)/2}$ is dimensionless. (a) Detector separation is always half the cavity size. (b) Detector separation is fixed relative to the cavity size. The zero mode impacts negativity up to some maximum length beyond which no entanglement can be extracted by the detectors regardless of the zero mode. The length $L$ is chosen so that the two detectors are very strongly spacelike separated despite the Gaussian tail of the switching function.}
            \label{fig: concur1}
        \end{figure*}
    
        \begin{figure*}[tp]
            \includegraphics[scale=0.8]{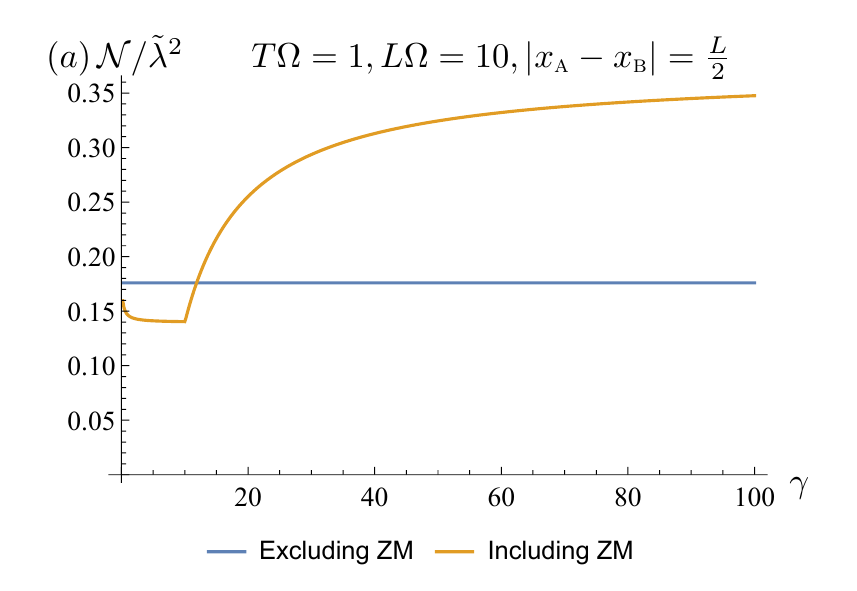}
            \includegraphics[scale=0.8]{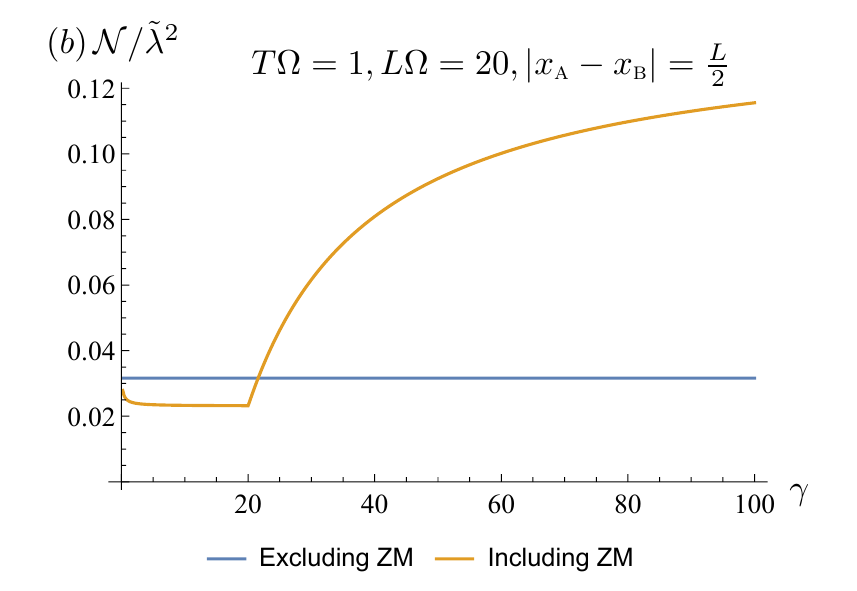}
            \includegraphics[scale=0.8]{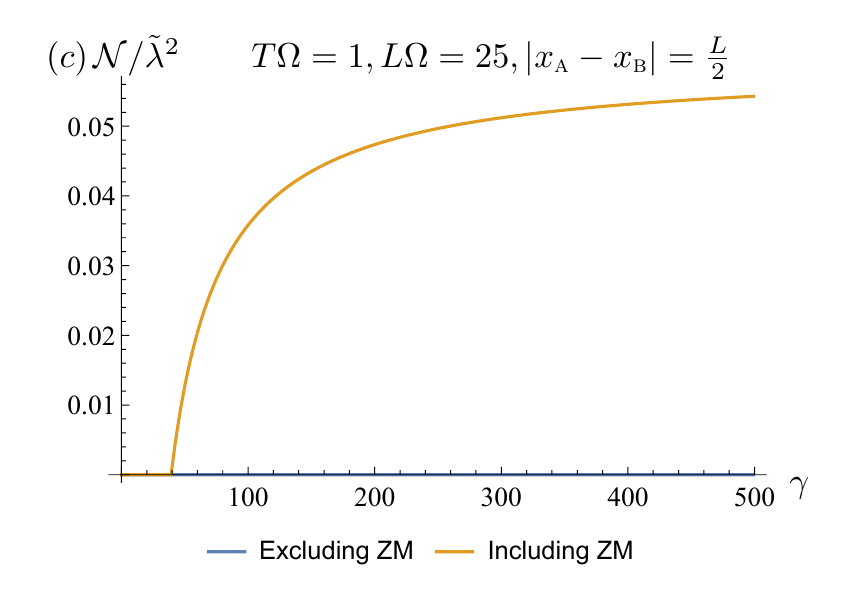}
            \includegraphics[scale=0.8]{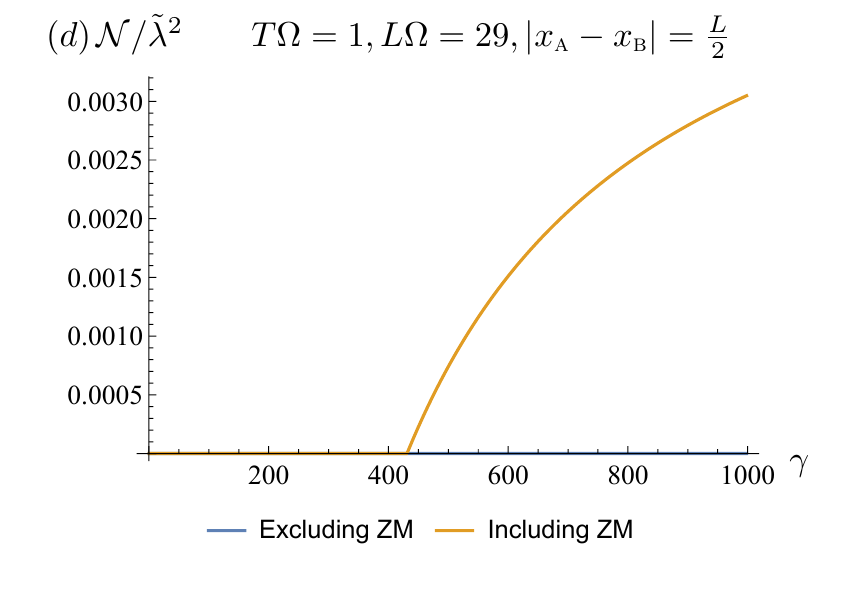}
            \caption{Negativity as a function of squeezing parameter of the harmonic oscillator wavepacket for the zero mode $\gamma$, for various choices of cavity size $L$ (in units of $\Omega$). In all plots the detectors are separated by half the cavity size ($|x_\textsc{a}-x_\textsc{b}|=L/2$). The interaction time is fixed at $T\Omega=1$. Here $\tilde\lambda = \lambda\Omega^{(n-3)/2}$ is dimensionless. For larger cavity (and correspondingly larger detector separation), the role of the zero mode is to amplify negativity at larger $\gamma$.}
            \label{fig: concur2}
        \end{figure*}
    
    In Figure~\ref{fig: concur1} and Figure~\ref{fig: concur2} we show the effects of the zero mode on the negativity of the two detectors for various choice of tunable setup parameters. The results clearly demonstrate how the zero mode can contribute significantly to the entanglement dynamics of the two detectors even at leading order in perturbation theory. Figure~\ref{fig: concur1} shows that for a larger cavity and larger detector separation, the zero mode leads to less efficient entanglement harvesting. Furthermore, Figure~\ref{fig: concur2} also shows that the amount of entanglement depends very much on the ``frequency'' of the harmonic oscillator wavepacket $\gamma$, which can be thought of as the squeezing parameter of the (Gaussian) state of the zero mode.  Note that for larger detector separation, the zero mode tends to improve entanglement generated by tuning the squeezing parameter $\gamma$ to be sufficiently large. This discussion illustrates strong dependence of entanglement harvesting on the choice of zero mode state, which is an ambiguity of the theory.

    
    \section{UDW-like models and generalization to higher dimensions}
    \label{sec: othermodels}
    
    In this section we briefly discuss the impact of the zero mode on entanglement harvesting for two classes of UDW-like models, namely 1) derivative coupling model, and 2) when the internal degrees of freedom of the detectors are described by harmonic oscillators. We will also comment on the generalizations to higher spacetime dimensions.

    \subsection{Derivative coupling model}
    
    For the derivative coupling model \cite{Louko2014firewall,Raval1996stochastic,Davies2002negativeenergy,Aubry2014derivative,Wang2014mirror}, the pullback of the Wightman function on the trajectory of the detector is
    \begin{align}
        \mathcal{A}(\sx(t),\sx(t')) &\coloneqq \tr\left(\hat\rho_\phi\partial_t\vp(\sx(t))\partial_{t'}\vp(\sx(t'))\right)\notag\\
        &= \partial_t\partial_{t'}W(\sx(t),\sx(t'))\,.
    \end{align}
    It is then easy to parallel what we did above and see how zero mode component affects the efficiency of entanglement harvesting for a derivative coupling. Using the expression for $C^\pm_\zm(t,t')$ in Eqs. \eqref{eq: hadzm} and \eqref{eq: comzm}, we can compute the derivative coupling Wightman function easily, which reduces to
    \begin{align}
        \mathcal{A}_\zm(\sx(t),\sx(t')) = \frac{\gamma}{2L^2}\,.
    \end{align}
    The relevant density matrix elements associated to zero mode are now
    \begin{align}
        \mathcal{L}^{\mathcal{A}}_\zm &= \lambda^2 \frac{\pi \gamma T^2 e^{-T^2 \Omega ^2}}{L^2}\,,\\
        \mathcal{M}^{\mathcal{A}}_\zm &= -\lambda^2\frac{\pi  \gamma  T^2 e^{-T^2 \Omega ^2}}{L^2}\,,
    \end{align}
    i.e. $\mathcal{M}^{\mathcal{A}}_\zm = -\mathcal{L}^{\mathcal{A}}_\zm$. Therefore, for derivative coupling, there are choices for the zero mode state such that its contribution to harvesting vanishes. Namely, for either one of the following limits: $T\to\infty$, $L\to\infty$, and $\gamma\to 0$ (compared to $T\to\infty$ or $\gamma\to\infty$ for linear coupling). Note that strictly speaking one cannot take the squeezing parameter $\gamma\to 0$ (or $\gamma\to\infty$) as it corresponds to non-normalizable eigenstate of $P_S$ (or $Q_S$). We also observe that unlike the linear coupling case, the derivative coupling zero mode contribution does vanish in the large cavity limit.
    
    Note that despite the fact that there are choices of the zero mode state that cancel the contribution to harvesting for the derivative coupling in some limits, it certainly does not mean that the derivative coupling is devoid of problems for harvesting in the presence of a zero mode. The derivative coupling has exactly the same issues as the usual UDW coupling. Concretely, the amount of entanglement harvested is still strongly dependent on the particular choice of zero mode state, and there is, in principle, no good reason to choose one state or another for the zero mode of the field in the Einstein cylinder. If one would like to choose a tailor-made state that makes the harvesting calculations coincide with the case of ignoring the zero mode, one would still need to justify why that choice of zero mode state is physical in any way.
    
    \subsection{Harmonic oscillator UDW detectors}
    
    We consider two pointlike harmonic oscillators of unit mass with frequencies $\omega_\textsc{A}$ and $\omega_\textsc{B}$ respectively. The interaction Hamiltonian is given by
    \begin{align}
        &\hat H_{I,\text{HO}}(t) \notag\\
        &= \sum_{\nu=\{\textsc{a,b}\}}\lambda_\nu\chi_\nu(t)\hat q_\nu(t)\int \d^n\bx \,F_\nu(\bx-\bx_\nu)\hat\phi(t,\bx)\,,
        \label{eq: hamiltonian-HO}
    \end{align}
    where $\nu=\text{A,B}$ labels the two detectors, and we couple to the quadrature
    \begin{align}
        \hat q_\nu(t) &= \hat{a}^{\phantom{\dagger}}_\nu e^{-\ii\omega_\nu t}+  \hat a^\dagger_\nu e^{\ii\omega_\nu t}\,,
    \end{align}
    with the  usual ladder operators of the harmonic oscillator $\hat a_\nu^{\phantom{\dagger}}$ and $\hat a_\nu^\dagger$ satisfying the commutation relation $[\hat a_\mu^{\phantom{\dagger}},\hat a_\nu^\dagger] = \delta_{\mu\nu}\openone$.  The ground state of each detector, denoted $\ket{0_\nu}$, is defined to be the state that satisfies $\hat a_\nu\ket{0_\nu} = 0$, where $\nu=\text{A,B}$. As in the two-level case in Section~\ref{sec: setup}, we consider the full ground state to be given by
    \begin{align}
        \hat\rho_0 =  \ket{0_\textsc{a}}\!\bra{0_\textsc{a}}\otimes\ket{0_\textsc{b}}\!\bra{0_\textsc{b}}\otimes\hat\rho_{\phi}\,,
    \end{align}
    where the field state is the same as before, i.e. $\hat\rho_\phi=\hat\rho_\zm\otimes\hat\rho_\osc$, with  $\hat\rho_\osc$ chosen to be the Fock vacuum, and the zero mode state is a harmonic oscillator ground state parametrized by $\gamma$ with first and second moments given by Eq.~\eqref{eq: moments}.
    
    Using the time evolution operator
    \begin{align}
        \hat U_{\text{HO}} = \mathcal{T}\exp\rr{-\int_{-\infty}^\infty\d t\,\hat H_{I,\text{HO}}(t)}\,,
    \end{align}
    the reduced joint density matrix of the detectors can be computed using 
    \begin{align}
        \hat\rho_{\textsc{ab}} = \tr_\phi\rr{\hat U_{\text{HO}}\hat\rho_0\hat U^\dagger_{\text{HO}}}\,.
    \end{align}
    At leading order in perturbation theory, the second excited state of each harmonic oscillator contribute to $\hat\rho_{\textsc{AB}}$: in terms of ordered basis $\left\{\ket{0_\textsc{a}0_\textsc{b}},\ket{0_\textsc{a}1_\textsc{b}},\ket{1_\textsc{a}0_\textsc{b}},
        \ket{1_\textsc{a}1_\textsc{b}},\ket{0_\textsc{a}2_\textsc{b}},\ket{2_\textsc{a}0_\textsc{b}}\right\}$, the leading order joint density matrix is a $6\times 6$ matrix given by \cite{hotta2020duality}
    \begin{align}
        \hat\rho_{\textsc{AB}} 
        &=
        \begin{pmatrix}
        1-\mathcal{L}_{\textsc{aa}}-\mathcal{L}_{\textsc{bb}} & 0 & 0 & \mathcal{M}^* & \mathcal{K}^*_\textsc{b} &\mathcal{K}^*_\textsc{a}\\
        0 & \mathcal{L}_{\textsc{bb}} & \mathcal{L}_{\textsc{ba}} & 0 & 0 & 0\\ 
        0 & \mathcal{L}_{\textsc{ab}} & \mathcal{L}_{\textsc{aa}} & 0 & 0 & 0\\
        \mathcal{M} & 0 & 0 & 0 & 0 & 0\\
        \mathcal{K}_\textsc{b} & 0 & 0 & 0 & 0 & 0\\
        \mathcal{K}_\textsc{a} & 0 & 0 & 0 & 0 & 0
        \end{pmatrix}\,,
    \end{align}
    where the $\mathcal{M},\mathcal{L}_{\mu\nu}$ are the same as Eq.~\eqref{eq: densitymatrixelements}. As discussed in earlier work \cite{hotta2020duality}, at leading order, there are new matrix elements that appear because we consider a harmonic oscillator detector rather than a qubit one, these are the terms $\mathcal{K}_\nu$ which take the form 
    \begin{align}
        \mathcal{K}_\nu &\coloneqq -\lambda^2_\nu\int_{-\infty}^\infty\d t\int_{-\infty}^t\d t'\,\chi_\nu(t)\chi_\nu(t')e^{\ii\omega_\nu (t+t')}\notag\\
        &\hspace{2cm} \times W(\sx_\nu(t),\sx_\nu(t'))\,,\hspace{0.5cm}\nu=\text{A,B}\,.
    \end{align}
     Again for simplicity we consider identical pointlike detectors ($\omega_\textsc{a}=\omega_\textsc{b}=\omega$, $\lambda_\textsc{a}=\lambda_\textsc{b}=\lambda$, $\chi_\textsc{a}(t)=\chi_\textsc{b}(t)=\chi(t)$) with smooth Gaussian switching given by Eq.~\eqref{eq: gaussian}, so the two negative eigenvalues of $\hat\rho_{\textsc{ab}}$ take the form
    \begin{subequations}
        \begin{align}
        E_1 &= -|\mathcal{M}| + \mathcal{L}_{\nu\nu} \,,\hspace{0.5cm} j =\text{A,B}\,,\\
        E_2 &= -|\mathcal{L}_{\textsc{ab}}|^2 - |\mathcal{K}_\textsc{a}|^2  - |\mathcal{K}_\textsc{b}|^2\,, 
    \end{align}
    \end{subequations}
    but since $E_2=O(\lambda^4)$, it does not affect negativity at leading order. Hence, to leading order we get \cite{hotta2020duality}
    \begin{align}
        \mathcal{N}_\text{HO} &= \max\{-E_1,0\} + O(\lambda^4) \notag\\
        &= \max\{|\mathcal{M}| - \mathcal{L}_{\nu\nu},0\} + O(\lambda^4) \,,
    \end{align}
    which is identical to Eq.~\eqref{eq: negativity} up to identification of the tunable parameters of each model, e.g. the two-level system's energy gap $\Omega$ with that of the harmonic oscillator $\Omega \to \omega$. This has also been pointed out in \cite{Brown2013harmonic,hotta2020duality}. Since the zero mode is unchanged, in the context of entanglement harvesting harmonic oscillator UDW model will share the same zero mode ambiguity problem as the two-level UDW model.
    
    We note that the formal expression for the density matrix is identical even if we consider derivative coupling, since only the Wightman function is changed from the amplitude coupling to derivative coupling case. Therefore, the computations done in the two-level case would carry through for harmonic oscillator case as well. In summary, the harmonic oscillator detector model presents the same zero-mode ambiguities as the qubit model both for derivative and amplitude coupling.
    
    \subsection{Generalizations to higher dimensions}
    
    The results obtained so far apply not only to the (1+1)-dimensional Einstein cylinder but also to any number of spatial dimensions. In particular, analogous results hold for a toroidal spacetime in $(n+1)$-dimensions where the spacelike hypersurface at constant $t$ has topology $S^1\times ...\times S^1$. The reason is that the zero mode is given by exactly the same expression except with the ``effective mass'' $L^n$ instead of $L$ in the zero mode Hamiltonian (see, e.g., \cite{Tjoa2019zeromode}):
    \begin{align}
        \hat H_{\zm,(n+1)} &= \frac{\hat P^2}{2L^n}\,.
    \end{align}
    Therefore, the features studied in this paper (and critically the strong dependence  of entanglement harvesting on the state of the zero mode)  will be present as well in arbitrary dimensions.
    

    \section{Conclusion}
    \label{sec: conclusion}
    
    We have discussed entanglement harvesting in spacetime topologies where a zero mode is present (e.g., the $(1+1)$-dimensional Einstein cylinder). In studying entanglement harvesting one couples two particle detectors to the field. The coupling of the detectors to the zero mode cannot be ignored. If the detectors do not couple to the zero mode they are able to signal superluminally \cite{Tjoa2019zeromode}, and as such, the study of spacelike entanglement harvesting becomes ill-defined unless the zero mode is considered.
    
    We have studied the role of the zero mode on  entanglement harvesting  with two different particle detector models (Unruh-DeWitt \cite{Unruh1979evaporation,DeWitt1979} and derivative coupling \cite{EMM2014zeromode,Louko2014firewall,Raval1996stochastic,Davies2002negativeenergy,Aubry2014derivative,Wang2014mirror}) that have been used in previous literature to study harvesting in the Einstein cylinder (e.g., \cite{Lin2016entangleCylin}). Due to the lack of Fock representation, for the zero mode we considered a family of Gaussian states parametrized by a `squeezing' parameter and showed that entanglement harvesting is strongly dependent on the choice of the zero mode state.
    
    There are two possible ways of reading this result: first, if one has control on the state of the zero mode one can see that the size of a periodic cavity,  and the detector separation affects entanglement harvesting through the zero mode dynamics. In particular, harvesting  is less efficient for a smaller cavity and  for larger detector separation. Furthermore, entanglement harvesting can be  improved  by tuning the state of the zero mode, for example, increasing the squeezing parameter for the family of Gaussian states considered in this paper. 
    
    Second, and perhaps the most important result: we observe that 1)---there are no \textit{a priori} reasons to choose one state or another for the zero mode (there is no `vacuum' state for the zero mode) and 2)---at leading order in perturbation theory, the zero mode state has a significant influence on the entanglement dynamics between two detectors. These two factors lead to the conclusion that the study of entanglement harvesting in spacetimes with periodic topologies suffers from strong ambiguities (again, in the absence of a good reason to select a state for the zero mode). This is true regardless of the nature of the detector-field coupling, e.g., whether it is a UDW coupling or a derivative coupling does not change this result.
    
    An interesting future direction is to consider how the entanglement harvesting protocol works in the compactified free boson model \cite{francesco2012conformal}. The compactified free boson model also has discrete spectrum, and it is the standard way of dealing with the zero mode in the context of two-dimensional CFTs that maintains causal behaviour and unitarity of the theory. It is reasonable to expect that the detector response and entanglement harvesting ambiguities can be solved in the compactified boson model. Comparing the results with the results in this paper is left for future work.

    \section*{Acknowledgment}
    E.T. acknowledges support from the Mike-Ophelia Lazaridis Fellowship. E. M-M. acknowledges support through the NSERC Discovery program and the Ontario Early Researcher Award.

\bibliography{zeromoderef}

\end{document}